\documentclass[runningheads]{llncs}

\usepackage[T1]{fontenc}
\usepackage{graphicx}
\usepackage{orcidlink}
\usepackage{bbding}

\begin{document}
\title{Decoding the Dashboard: Data Comics to Support Students’ Understanding of Learning Analytics Visualisations}
\titlerunning{Decoding the Dashboard}

\author{
    Mikaela Elizabeth Milesi\inst{1}\textsuperscript{(}\Envelope\textsuperscript{)}\orcidlink{0009-0002-0910-9822}
    \and
    Vanessa Echeverria\inst{2, 3}\orcidlink{0000-0002-2022-9588}
    \and
    Lixiang Yan\inst{4}\orcidlink{0000-0003-3818-045X}
    \and
    Yueqiao Jin\inst{1}\orcidlink{0009-0003-7309-4984}
    \and
    Riordan Alfredo\inst{1}\orcidlink{0000-0001-5440-6143}
    \and
    Jie Xiang Fan\inst{1}\orcidlink{0009-0000-8585-2760}
    \and
    Linxuan Zhao\inst{1}\orcidlink{0000-0001-5564-0185}
    \and
    Dragan Gašević\inst{1}\orcidlink{0000-0001-9265-1908}
    \and
    Yi-Shan Tsai\inst{1}\orcidlink{0000-0001-8967-5327}
    \and
    Roberto Martinez-Maldonado\inst{1}\orcidlink{0000-0002-8375-1816}
}

\authorrunning{M. Milesi et al.}

\institute{Faculty of Information Technology, Monash University, 25 Exhibition Walk, Clayton 3168, Victoria, Australia \email{\{mikaela.milesi, ariel.jin, riordan.alfredo, jie.fan, linxuan.zhao, dragan.gasevic, yi-shan.tsai, roberto.martinezmaldonado\}@monash.edu} \and
School of Computing Technologies, RMIT University, Melbourne 3000, Victoria, Australia \email{vanessa.echeverria@rmit.edu.au} \and
Escuela Superior Politécnica del Litoral, ESPOL, Guayas, Ecuador \and
Tsinghua University, Beijing, China \email{lixiangyan@mail.tsinghua.edu.cn}
}

\authorrunning{Milesi et al.}

\maketitle
\begin{abstract}
Learning analytics dashboards (LADs) are intended to help students make sense of their learning data to support reflection and decision-making. However, their visualisations can be complex, particularly for students with low visualisation literacy. Narrative techniques, such as annotated charts and data comics, have been used to communicate insights directly, but not as supplementary materials to empower students to explore their visualisations themselves. In response, we conducted a qualitative study examining how data comics can complement LADs. We interviewed 18 nursing students and 4 of their teachers about a multimodal LAD containing visualisations with data comics explaining them. Analysis showed that data comics were clear, engaging, and helped make complex visualisations more accessible, though they must be carefully designed to avoid overwhelming students with information. The findings suggest that both students and teachers are receptive to data comics as a means of supporting the interpretability of LADs.

\keywords{data comics \and information visualisation \and learning analytics \and dashboards}

\end{abstract}
\section{Introduction and Background}
\vspace{-2mm}
Learning analytics dashboards (LADs) are tools designed to help students and teachers reflect on learning experiences and make data-informed decisions \cite{greller2012translating}. Defined by Schwendimann et al.~\cite[p.37]{schwendimann2017perceiving} as \textit{``a single display that aggregates different indicators about learner(s), learning process(es), and/or learning context(s) into one or more visualisations''}, the notion of the dashboard has been widely appropriated within the field of LA to support the development of analytical thinking in students \cite{rudenko2025dashboard}, help students engage with feedback \cite{alcock2024visualisations}, and prompt critical reflection \cite{echeverria2025learning}. LADs incorporate a range of visualisations \cite{sahin2021visualizations}, from familiar graphs \cite{Park2019}, such as bar or line charts, to potentially complex visualisations of multimodal LA data \cite{zhao2024towards}.

A major challenge associated with LADs is that students may not possess the necessary data skills \cite{fernandeznieto2021storytelling,fernandeznieto2022beyond} or \textit{visualisation literacy} \cite{donohoe2020data,yan2025effects} to effectively derive insights from LA visualisations unassisted. To address this, several studies \cite{echeverria2018exploratory,fernandeznieto2024editor} have proposed transitioning from an \textit{exploratory} approach -- where individuals are expected to interpret visualisations themselves and extract meaning \cite{knaflic2015storytelling} -- to an \textit{explanatory} approach where visualisations are enhanced with narrative elements (both textual and visual) to communicate insights directly \cite{knaflic2015storytelling}. Yet, there are still criticisms associated with this approach. Empirical studies have shown that the addition of narrative elements does not guarantee increased engagement with the information \cite{echeverria2018exploratory,boy2015storytelling,pozdniakov2024investigating}. Teachers from a study conducted by Echeverria et al.~\cite{echeverria2018exploratory} felt that narrative elements could impact engagement by adding visual noise, complexity, or redundant information to the visualisations. Similarly, Pozdniakov et al.~\cite{pozdniakov2024investigating} observed that the addition of narrative elements to LADs has a minimal effect on task completion accuracy. Furthermore, by communicating insights directly, narrative explanations may discourage individuals exploring visualisations themselves to formulate their own insights or opinions \cite{milesi2025piecing}. These criticisms highlight the need for further research on how narrative explanations can be integrated into LADs in ways that support, rather than replace, students’ own exploration and sensemaking.

In human–computer interaction (HCI) literature, data comics have emerged as a visualisation genre to encourage individuals to develop visualisation literacy skills and better engage with data \cite{wang2019teaching,boucher2023edudatacomics}. Inspired by traditional comics \cite{segel2010narrative,mccloud1994understandingcomics}, data comics communicate data-driven information through a combination of illustrative and textual content, the organisation of panels, and representations of data (ranging from conventional charts to abstract depictions) \cite{bach2017emerging}. Despite research suggesting that data comics are an appropriate format to communicate complex information to non-scientific audiences \cite{bach2018designpatterns,wang2021data,liang2024data}, there is limited work examining their effectiveness in supporting students’ sensemaking of learning analytics data. To the best of our knowledge, the only existing study using data comics in the context of LA indicates that, while data comics can increase engagement, their impact varies across learner groups and they are often better suited as supplementary rather than primary instructional resources \cite{milesi2026comicpanels}. This points to a key gap: it remains unclear how narrative visual formats such as data comics can be designed to support students’ active exploration and interpretation of LADs.

This paper addresses this gap by presenting a qualitative study, situated within an authentic educational setting, where 18 nursing students and 4 teachers were asked to comment on whether data comics facilitate understanding of multimodal LA visualisations, with teachers included as key stakeholders in the design and pedagogical integration of such tools, as suggested by Kaliisa et al. \cite{kaliisa2023checklist}. This study aimed to provide a preliminary exploration into the potential for data comics to support interpretation of LA visualisations through the following research questions:
\vspace{-3pt}
\begin{itemize}
    \item \textbf{RQ1:} How do \textit{students}, considering differences in their \textit{visualisation literacy}, perceive data comics as a supplementary tool to support the interpretation of student-facing learning analytics visualisations?
    \item \textbf{RQ2:} How do \textit{teachers} perceive data comics as a supplementary tool to explain student-facing learning analytics visualisations?
\end{itemize}

\section{Study Design}
\subsection{Learning Context}
This study is situated in an authentic learning scenario designed to develop teamwork and communication skills in undergraduate nursing students. The clinical simulations, created by teachers at Monash University, required teams of four students to collaboratively manage the care of four manikin patients.

Each nursing student was equipped with a headset microphone to record audio data, a \textit{Fitbit Sense} smart watch to collect their physiological data (i.e., heart rate), and a Pozyx position tag to capture their location (in the form of $x$-$y$ coordinates) and body orientation within the ward.

\begin{figure}[h]
    \centering
    \includegraphics[width=0.95\linewidth]{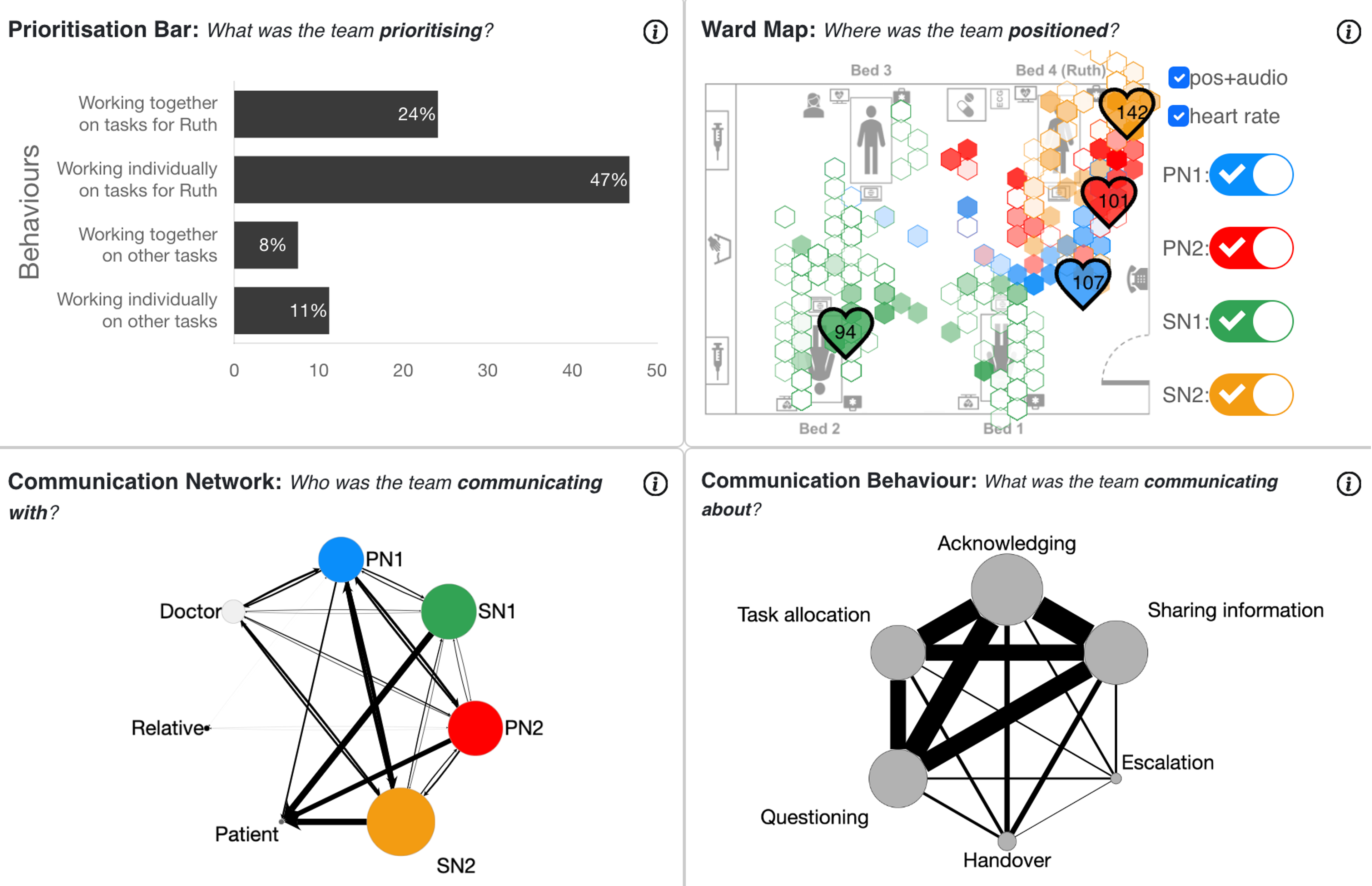}
     \vspace{-4mm}
    \caption{Multimodal visualisations in the LAD. The \textit{Prioritisation Bar} chart represents the individual and team-based behaviours that students were engaging in during the simulation. The \textit{Ward Map} depicts where the students were positioned, the extent to which they were talking and the maximum heart rate in relation to the learning space. The \textit{Communication Network} shows who each role in the simulation were speaking to via arrows and how often they were speaking, as indicated by the size of each node. The \textit{Communication Behaviour} network visualises what students were discussing categorised into six key communication behaviours.}
    \label{fig:mmla-vis}
    \vspace{-10mm}
\end{figure}

The clinical simulation was closely followed by a debriefing session whereby students reflections were guided by a member of the teaching staff while they viewed a LAD, as a part of their regular classroom activities, that was co-designed with teachers and validated by Echeverria et al. \cite{echeverria2025teamvision}. The LAD displayed the following four visualisations of the multimodal data that depict teamwork and communication behaviours during the simulation (Figure \ref{fig:mmla-vis}):

\noindent\textbf{Prioritisation bar:} A bar chart that depicts the amount of time that students engaged in particular behaviours during the simulation based on their positioning data. These behaviours relate to tasks that are completed for the deteriorating patient or other clinical tasks. The behaviours are also categorised based on whether it was a task performed by an individual or more than one student (team task).
\textbf{Ward map:} A hexbin map that combines the audio, positioning and heart rate data to represent where students were located within the simulation ward, as well as how often they were speaking in that position. The colours of the hexagons indicate the roles of the students in the simulation (i.e., primary nurse 1 is blue, primary nurse 2 is red, secondary nurse 1 is green, and secondary nurse 2 is orange). The opacity of the hexagon represents the amount of time spent speaking (i.e., a more intense colour indicates more time communicating). The heart-shaped icons with a number inside represent where the maximum heart rate was detected for each student.
\textbf{Communication network:} A sociogram that uses the audio data to show how each role in the simulation communicated with each other. The arrows indicate the directionality of the communication, with the thickness of the arrow representing how often two roles in the simulation were talking to each other. The size of the circle shows how often an individual spoke over the simulation.
\textbf{Communication behaviour:} An epistemic network that shows key communication behaviours in healthcare, as identified by teachers in the study by Zhao et al. \cite{zhao2024towards}. The audio data was transcribed using the OpenAI Whisper medium model \cite{radford2023robust} and coded using a fine-tuned BERT-based classifier \cite{zhao2024towards} as: 1) task allocation, 2) questioning, 3) handover, 4) escalation, 5) sharing information, and 6) acknowledging. 

\subsection{Participants and Procedure}
The participants of this study were a subset of students who participated in the clinical simulation, as well as members of the teaching team who facilitated the simulation and debriefing session. A total of 22 participants, 18 nursing students (S1-S18; 15 females, 3 males; age between 20 and 49 - $\mu$ = 24.2, $\sigma$ = 6.5) and 4 teachers (T1-T4; 4 females), volunteered to be interviewed as part of this study. Students received a \$50 AUD voucher in compensation for their participation.

Ethics approval was obtained from the Monash University Human Research Ethics Committee (Project ID: 39190). Informed consent was secured from each participant prior to the interview. No additional personal information was collected as part of this study.

\textbf{Mini-VLAT questionnaire:} Before the interviews, students completed the Mini-VLAT \cite{pandey2023minivlat} through \textit{Qualtrics} to assess their visualisation literacy. Since this study examines the role of data comics in supporting the interpretation of multimodal LA visualisations, measuring visualisation literacy helped determine the extent to which data comics were beneficial for students with varying levels of proficiency. The ``correction-for-guessing'' formula \cite{frary1988formula} was used to calculate a corrected Mini-VLAT score. Additionally, participants were given an option \textit{``I'm not sure''} to discourage guessing the answer.

The interviews were conducted via Zoom. Students participated in a 75-minute long interview, as they were asked to reflect on their own data. In contrast, teachers took part in shorter 45-minute long interviews, focusing primarily on feedback on the potential benefits and criticisms at a higher level. The interviews included the following tasks:

\noindent\textbf{Task 1 -- Exploration of the LAD:} Students freely explored the dashboard and visualisations generated from their own simulation data. In contrast, the teachers were shown data from a single session that was purposefully chosen as the visualisations of the multimodal data were representative of the teamwork and communication interactions that were typically observed across most of the teams. Each visualisation featured an information icon, \textcircled{\textit{i}}, which opened a data comic explaining the visualisation when clicked. Participants were encouraged to spend $\sim10$ minutes interacting with the LAD to make sense of the visualisations and ask any clarification questions to the interviewer.

\noindent\textbf{Task 2 -- Evaluation of Data Comics:} Participants were shown each data comic in the LAD and asked to describe the extent to which it supported interpretation of the visualisations, focusing on how data comics helped students make sense of their own data (RQ1) and whether they supported teachers in explaining students’ visualisations (RQ2), along with its perceived benefits and limitations as a supplementary explanatory tool.

The interviews were recorded via Zoom and transcribed using the OpenAI Whisper (large version) and speaker diarisation model. An inductive thematic analysis \cite{braun2006thematicanalysis} was conducted and discussed by three researchers until a consensus was reached. 
\vspace{-2mm}
\section{Data Comics Design}
\begin{figure}[htbp]
    \centering
    \includegraphics[width=1.00\linewidth]{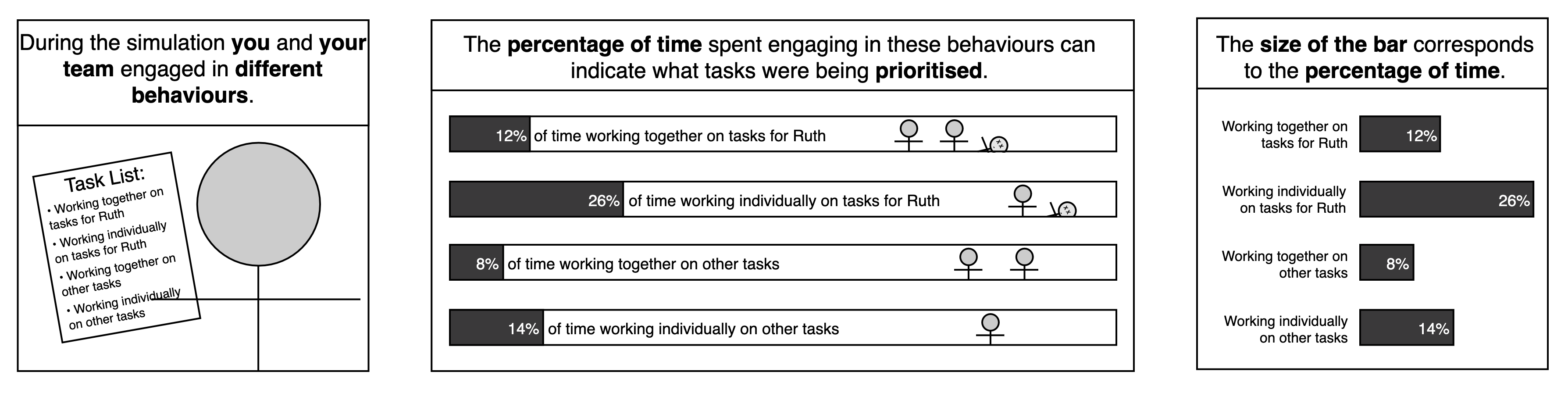}
    \caption{Data comic to explain the \textit{prioritisation bar} visualisation with three panels: the first explains the context behind the visualisation (i.e., the different behaviours), the second introduces percentages to indicate that the size of the bar corresponds to the amount of time spent engaging in a particular behaviour, and the last panel removes any stylistic elements in favour of showing the bar chart.}
    \label{fig:prioritisation}
\end{figure}

There were four data comics prototypes developed for this study -- one for each visualisation in Figure \ref{fig:mmla-vis}. Each data comic uses a \textit{linear} panel layout \cite{bach2018designpatterns} and follows a left-to-right reading order, continuing top-to-bottom when there is more than one row \cite{cohn2013navigating} (e.g., see Figure \ref{fig:prioritisation}).

\begin{figure}[htbp]
    \centering
    \includegraphics[width=1.00\linewidth]{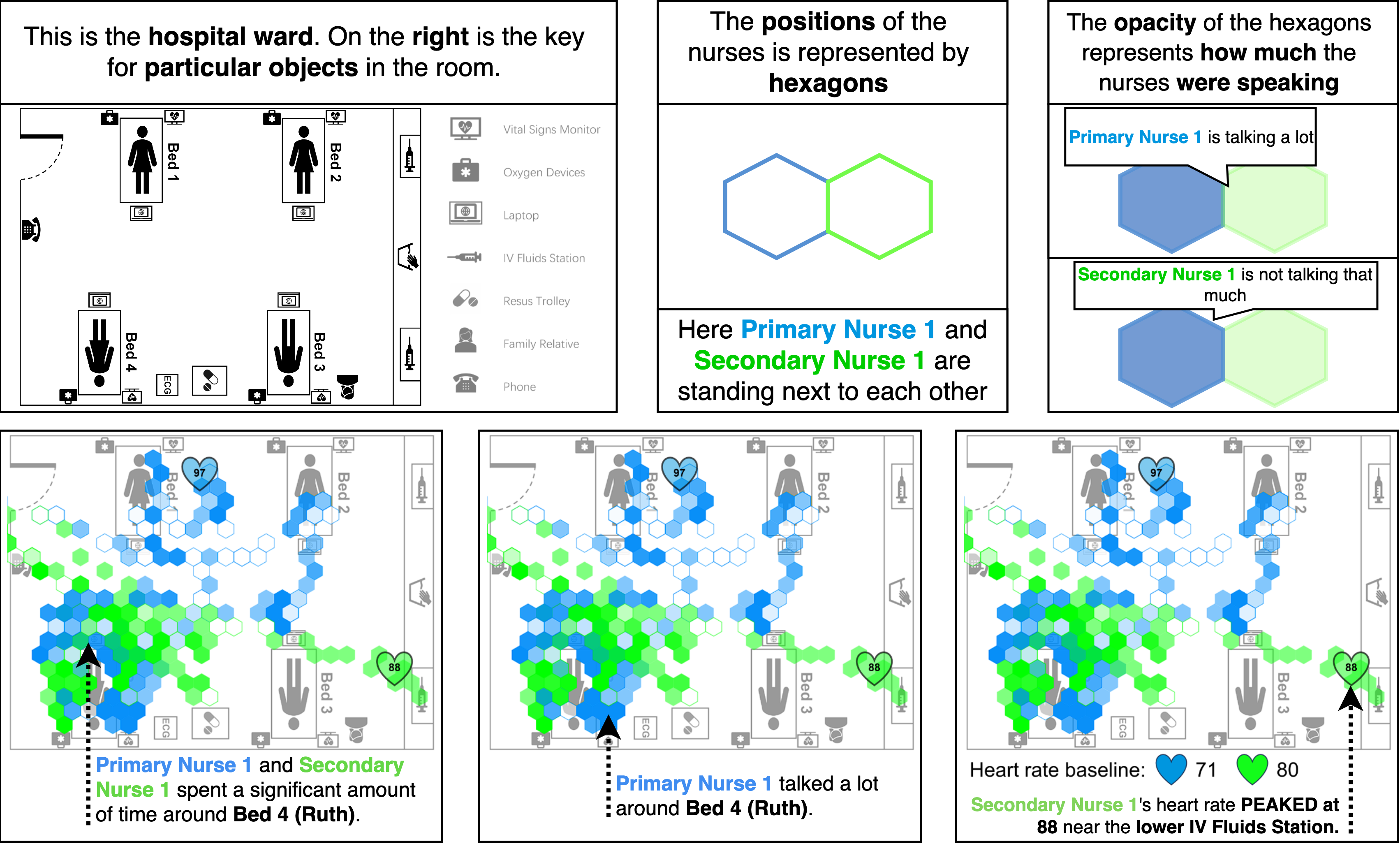}
    \caption{Data comic to explain the \textit{ward map} visualisation. There are six panels each containing the ward map visualisation and a textual annotation of salient information.}
    
    
    
    
    
    \label{fig:ward-map}
\end{figure}

\begin{figure}[htbp]
    \centering
    \includegraphics[width=1.0\linewidth]{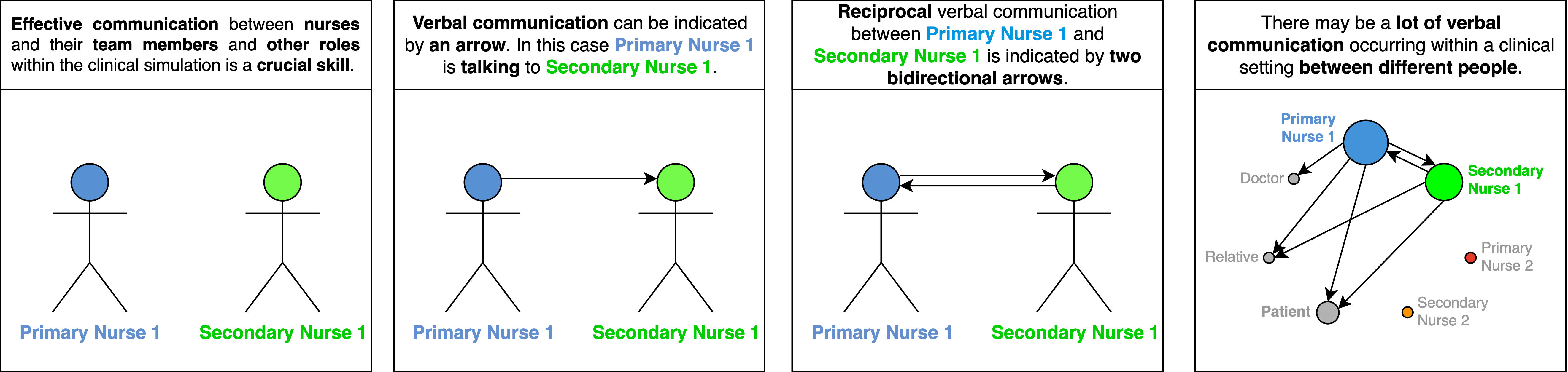}
    \caption{Partial data comic to explain the \textit{communication network} visualisation. This samples the first four panels which show what the arrows are indicating.}
    \label{fig:comm-network}
\end{figure}

\begin{figure}[htbp]
    \centering
    \includegraphics[width=1.0\linewidth]{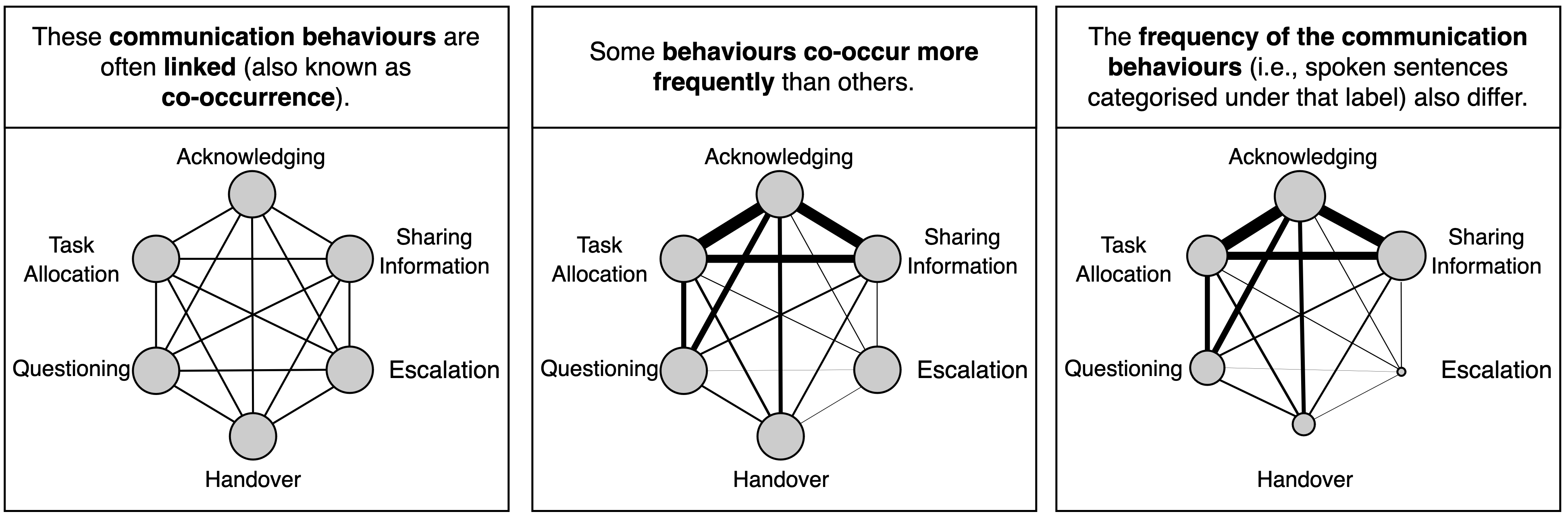}
    \caption{Partial data comic to explain the \textit{communication network} visualisation. This samples the last three panels that explain the concept of co-occurrence.}
    \label{fig:comm-behaviour}
\end{figure}

An abstract form of the \textit{visualisation build-up} pattern was also used for each prototype, where each panel adds information needed to interpret the visualisation as a whole. Unlike conventional chart build-ups (e.g., adding axes or legends), these data comics begin with a stylised representation of the information using traditional comic elements like stick figures and speech bubbles before gradually transitioning into the respective multimodal LA visualisation. Each panel combines text and imagery: text provides the necessary information, while images illustrate what is described \cite{mccloud2006makingcomics}. They are designed to communicate one message per panel \cite{bach2018designpatterns}. This paper presents the full versions of the \textit{prioritisation bar} (Figure \ref{fig:prioritisation}) and \textit{ward map} (Figure \ref{fig:ward-map}) data comics, and partial versions of the \textit{communication network} (Figure \ref{fig:comm-network}) and \textit{communication behaviour} (Figure \ref{fig:comm-behaviour}) data comics (\href{https://anonymous.4open.science/r/DataComics-NursingSuppMaterial-8340/}{full versions can be accessed here}).

\vspace{-10pt}
\section{Results}
Students achieved an average corrected Mini-VLAT score of 8.26 out of 12 ($\sigma=1.79$). The scores ranged from 3.33 to 10.67 (median = 8.67). For the purposes of this study, we refer to a score of less than 8 as being low relative to the cohort (n = 6), 8-9 as being mid-range (n = 7), and greater than 9 as being relatively high (n = 5). 

This resulting themes and codes from the thematic analysis can be seen in Table \ref{tab:thematic}. 
\begin{table}[htbp]
    \centering
    \caption{Themes emerging from the inductive thematic analysis}
    \begin{tabular}{|l|p{4cm}|p{4.25cm}|}
        \hline
        \textbf{Theme} & \textbf{Emerging Codes} & \textbf{Description}\\ \hline
        Benefits of Data Comics & Clarify information; Familiar format; Ease of interpretation; Engaging; Unique; Aid in recall & Participants spoke positively about features of the data comics or highlighting situations where they could be beneficial for learning.\\ \hline
        Criticisms of Data Comics & Visualisations are self-explanatory; Recognised visualisations from debriefing; Information overload & Participants criticised the data comics or identified factors that would render the data comics unnecessary.\\ \hline
        Design Suggestions & Use of colour or visual features; Character design; Comic content & Participants commented on ways that the data comic prototypes could be improved.\\ \hline
    \end{tabular}
    \label{tab:thematic}
\end{table}

\subsection{Data comics aid interpretation}
Both students and teachers perceived data comics as useful supplementary tools to clarify information that was not obvious to them previously or make the process of interpreting the student-facing learning analytics visualisations easier.

Regarding \textbf{RQ1}, students across all visualisation literacy levels identified that the data comics made the multimodal LA visualisations \textit{``easier to understand''} (S1, 7.00; S15, 9.33) and \textit{``make much more sense''} (S3, 8.00). 

Students with lower Mini-VLAT scores praised the \textit{``simple''} design that allow them to \textit{``check that [they are] interpreting [the visualisations] correctly''} (S1, Mini-VLAT score = 7.00). This sentiment was shared by S7 (7.33) who stated, \textit{``if you didn't [know] what the graphs mean then [the data comics] tells you''}. These students also mentioned that data comics provided valuable support and context to enable them to interpret the visualisations. For example: S18 (Mini-VLAT score = 6.00) said: \textit{``most graphs I see have no context whenever I'm researching, and researching is something I find quite hard anyway, so the fact that I can explain it in a comic sense, and I am a visual learner in general''}. This was echoed by S10 (7.00) who said: \textit{``it gives you more information if you need it … If you need to understand the tool a bit better, then you can go in and read more''}. 

In contrast, those students with relatively high Mini-VLAT scores related the utility of the data comics to the complexity of the visualisations stating: \textit{``there's a lot going on in the [visualisations] so it takes a lot to explain it. The nature of the graphs are [more] complex than normal''} (S13, 10.67) and \textit{``the comics are really handy because there is a lot going on in these visualisations. … It's easy for anyone to understand and there's a lot of information because otherwise I think [the visualisations] could get a little bit confusing, but I know they're very helpful''} (S11, 10.67). S15 (9.33) believed the data comics allowed them to \textit{``tap into the smaller bits of information of how to interpret the graphs''}.

Regarding \textbf{RQ2}, The teachers expressed similar sentiments, stating that the data comics \textit{``provide context to students … We think it's obvious, but that's because we know the data''} (T3) and \textit{``I love that [students] can have a little description of what [the visualisation] means''} (T4). 

\subsection{Engaging format}
Beyond the explanatory role that data comics can have in LADs, students and teachers identified that the comic book style contributed to personal engagement and perceived ease of interpretation. Regarding \textbf{RQ1}, a student with a mid-range Mini-VLAT (S16; 8.67) commented, \textit{``It's a really easy format to read … because it's just little squares … you can follow along. … There's not too much going on. It really emphasised one point and one point only. It's just very reader-friendly''}. Also, a student with a mid-range Mini-VLAT score (S16, 8.67) highlighted the visual features of the \textit{ward map} data comic as being helpful: \textit{``It's a simple image and it also gives you a very brief description of all the components of the map. It's nice because it's colour coded, you have bolding on [the] words. So that makes it easier to digest.''}

Regarding \textbf{RQ2}, most teachers felt that the data comics were an engaging and unique way to present information. T1 described the data comics as \textit{``much more visually appealing''}, adding that they \textit{``could potentially improve engagement instead of just a slab of information [and be] beneficial for both … people who are visual learners or [those who] concentrate more on … text''}. T2 believed that \textit{``it doesn't stand out as a comic to be honest it stands out as something different. I like that … it's not sort of academic in a sense''}. T4 concluded that \textit{``students would prefer [data comics]''} over textual explanations. 

One of the teachers (T2) described the data comics as \textit{``more digestible. [When] the information [is] in that format you don't need to then work out - what are the figures? It's literally text and it seems concise and it's not overloading''}. This view was shared by students with relatively high Mini-VLAT scores (RQ1), such as S11 (10.67), who when describing the \textit{communication network} data comic explained, \textit{``I like how they're presented. I didn't notice before how it's the two people and then it's the arrows and then they take the people away. That makes a lot of sense [and are] definitely helpful''}. S13 (10.67) simply stated, \textit{``I think they're fun''}.

\subsection{Perceived redundancy and information overload}
\vspace{-6pt}
Regarding \textbf{RQ1}, although data comics were generally perceived as beneficial, some students considered them unnecessary. The perception that the visualisations were self-explanatory emerged from some students across visualisation literacy levels. For instance, S12 (7.67) noted that: \textit{``the graphs were quite straightforward''}, while S15 (9.33) described them as \textit{``easy to interpret … and understand''}. Familiarity also played a role: as the study was embedded in regular classroom use of the LAD, several students had already encountered the visualisations during prior debriefing sessions. Students with mid-range Mini-VLAT scores reported that they \textit{“could understand [the visualisations] without [the data comics] because we went over them in class … and it’s not too difficult to understand”} (S14, 8.67), and \textit{“I was counting on my teachers … during the debrief. I felt like I had a good idea of what was what”} (S18, 8.33). Yet, even students with higher Mini-VLAT scores acknowledged the value of data comics, particularly for first-time interpretation. S9 (9.33) stated that \textit{“if I was looking at it for the first time, I would definitely use the [data comics]”}, while S4 (10.67) highlighted how the comics clarified previously overlooked aspects of the visualisations, such as the co-occurrence of interactions.

The potential for information overload was also identified as a limitation of the data comic format. A student with a relatively low Mini-VLAT score (S10, 7.00) criticised the volume of text, stating, \textit{“I didn't read it all. … There's a lot of information”}. This concern was echoed by a student with a mid-range Mini-VLAT score, S8 (8.33), who felt that \textit{“I just need to understand the different connotations of [the visualisations] … The [data comic] doesn't need to be this complex”}. S8 further noted non-linear reading patterns, explaining they might \textit{“not necessarily read [the panels] from left to right”}, instead treating them as independent explanations.

Regarding \textbf{RQ2}, The teachers expressed similar concerns. When reflecting on the \textit{ward map} data comic T1 noted that, \textit{``It's pretty full on. I wonder if it could be condensed to three boxes''}. T2 worried about the density of information, describing the data comics as \textit{``very busy''}. On the topic of \textit{``cognitive load''} they questioned, \textit{``How do we engage them [within] a fifteen second window [without students] thinking, `Nah, can't be bothered'?''}. T2 suggested reducing the number of panels to \textit{``three boxes … start, middle, end … [with] a quicker progression''}. They cautioned that, \textit{``if you're going to give the information, it needs to be something additional that they wouldn't have picked up on''}.

\section{Discussion}
\subsection{Interpreting multimodal LA visualisations with data comics}
This study investigated how students with differences in visualisation literacy levels (RQ1) and their teachers (RQ2) perceived data comics as a supplementary tool to explain student-facing learning analytics visualisations. The comments from students and teachers show that the key strengths associated with adding data comics that explain LA visualisations to LADs are: i) they make LA visualisations easier to understand by providing additional context, ii) they simplify LA data to students in a way that supports interpretation and iii) they are quick and easy to read. This result broadly supports the findings of other studies in LA linking the implementation of narrative elements in LADs to improved sensemaking among non-data experts by providing contextual information \cite{fernandeznieto2021storytelling}. The results from that study and the current work highlight the value of narrative explanations for communicating LA data to students. However, our findings imply that narrative explanations can empower students to explore LA visualisations by guiding them through the interpretation process, rather than conveying insights to them directly. Yet, further research is needed to compare narrative elements that guide students through interpretation of LA visualisations with those that communicate insights.

Notably, benefits were identified by students across levels of visualisation literacy. However, where participants with relatively low Mini-VLAT scores expressed these benefits in terms of themselves and their own abilities to interpret the visualisations correctly, students with relatively high Mini-VLAT scores were able to relate these benefits to how the data comics enabled to navigate through the complexities and/or nuances of the data. In a quantitative study, Milesi et al.~\cite{milesi2025piecing} found that the addition of narrative elements benefited individuals in terms of supporting interpretation, yet warned that underlying messages from the data could be missed if narrative elements are not applied thoughtfully. These findings indicate that the implementation of narrative explanations can support a diverse range of users in both non-educational and learning contexts. However, these results also suggest a need for narrative elements to support exploratory data analysis rather than acting purely as explanatory mechanisms.

Kaliisa et al.~\cite{kaliisa2023checklist} stated that a key question in LAD design is how to engage users. In this study, teachers highlighted the challenges involved with engaging students quickly and felt that data comics could be more effective than text in this regard. Data comics have been recognised as an engaging medium to communicate information about visualisations in HCI literature \cite{wang2019comparing,bach2018designpatterns}; therefore, these findings suggest that data comics could potentially address issues relating to engagement with student-facing LA tools.

Some students across visualisation literacy levels did not see the value in data comics to explain LA visualisations, feeling that the visualisations were self-explanatory. This is consistent with teachers with high visualisation literacy from Pozdniakov et al.~\cite{pozdniakov2024investigating} who also found the narrative features to be redundant. While it could be assumed that data comics as supplementary tools are unnecessary for learners in contexts where visualisation literacy skills are necessary (e.g., STEM), one unanticipated finding was that even a student with a mid-range Mini-VLAT score struggled to interpret the layout/reading order of the data comic itself, showing that even students with moderate visualisation literacy skills can encounter difficulties when interpreting representations of data-driven information. Therefore, a takeaway message from this is that interfaces that include these narrative-enriched supplementary tools should include interactivity so that these features remain optional for those with high visualisation literacy, while still supporting students with lower visualisation literacy (or those who require assistance) in interpreting the LA visualisations.

The students and teachers both identified \textit{information overload} as being a primary issue with the data comics. According to comics theorist Scott McCloud \cite{mccloud1994understandingcomics}, a crucial skill for comic creators is to balance information in line with their assumptions about the readers. In our study, the teachers felt that students should have been able to extrapolate on information by drawing on their knowledge of the clinical simulation. This concept of \textit{decluttering} visualisations to remove visual features with minimal informative value has also gained traction in LA literature as a way to direct focus and attention towards salient information in annotated towards \cite{echeverria2018exploratory,pozdniakov2022question}. This idea can be extended to data comics by critically evaluating whether each individual panel and its content need to be decluttered or removed based on the students and learning context, alongside "traditional" graphical features of data visualisations like data points, borders, or headings \cite{knaflic2015storytelling}

\subsection{Implications for Educational Design and Practice}
Based on the comments from students and teachers, we formulated the following recommendations for LAD designers:

\noindent1) To engage students to explore LADs, data comics should be used to explain how to interpret LA visualisations over verbose text-based alternatives.

\noindent2) Data comics should be limited to showing only the essential information that is required to interpret LA visualisations. If implemented in a LAD, the interactive nature could be leveraged to offer a more detailed version of the data comic if the student requires additional assistance.

\noindent3) As the majority of design suggestions came from teachers, they should be involved in the process of designing the data comics to ensure they are contextually relevant for the learning activity, as well as appropriate (in terms of factors like content, information density, length and pedagogical goals) for their students.

These results present the potential that data comics have as a way to empower students to interpret LA visualisations and engage them with their learning data.

\subsection{Limitations \& Future Works}
This study is qualitative and exploratory in nature, and hence the sample size is small. While not meant for generalisation, the findings captured insightful views of educators and students in a particular clinical simulation learning context towards the use of data comics for learning reflections that may be tested further with a larger and more diverse sample. It is worth noting that the participants had already been exposed to the visualisations, so data comics were not a first encounter and may have been perceived as unnecessary. Yet, the study was non-experimental, embedded within an LAD already integrated into mandatory debriefs—one of the first multimodal LA interventions tested under fully authentic conditions. While this enhances ecological validity, future research should involve more diverse participants across disciplines and settings, and adopt controlled experimental designs to compare data comics with text-based narrations, especially in first-encounter scenarios.
\vspace{-3mm}
\section{Concluding Remarks}
\vspace{-2mm}
This paper examines the use of data comics as a supplementary tool to help students interpret LA visualisations. We interviewed students and teachers about comics depicting multimodal LA data from an authentic scenario. Results indicate that data comics are clear, engaging, and aid interpretation of complex visualisations regardless of visualisation literacy, though careful design is required to prevent overload. Both groups preferred data comics over text-based explanations for supporting their understanding. These findings suggest promising future applications of data comics in LA for explaining visualisations and communicating insights.

\bibliographystyle{splncs04}
\bibliography{bibliography}

\end{document}